\documentclass[sigconf]{acmart}
\renewcommand\footnotetextcopyrightpermission[1]{} 
\setcopyright{none} 
\pdfoutput=1    
\usepackage{graphicx}  
\usepackage{caption}   
\usepackage{subcaption}  
\usepackage{indentfirst}
\usepackage{booktabs}
\usepackage{tabularx}
\usepackage{fancyhdr}   
\AtBeginDocument{%
  }

\begin{document}
\title{Discriminating Human-authored from ChatGPT-Generated Code Via Discernable Feature Analysis}
\author{Ke Li}
\authornote{Both authors contributed equally to this research.}
\email{luck_ke@hust.edu.cn}
\author{Sheng Hong}
\authornotemark[1]
\email{hongsheng@hust.edu.cn}
\affiliation{%
  \institution{School of Cyber Science and Engineering, Huazhong University of Science and Technology, Wuhan, Hubei, China}
  \streetaddress{1037 Luoyu Road}
  \city{Wuhan}
  \state{Hubei}
  \country{China}
  \postcode{430074}
}

\author{Cai Fu}
\email{fucai@hust.edu.cn}
\affiliation{%
  \institution{School of Cyber Science and Engineering, Huazhong University of Science and Technology, Wuhan, Hubei, China}
  \streetaddress{1037 Luoyu Road}
  \city{Wuhan}
  \state{Hubei}
  \country{China}
}

\author{Yunhe Zhang}
\email{yhzhang@hust.edu.cn}
\affiliation{%
  \institution{School of Cyber Science and Engineering, Huazhong University of Science and Technology, Wuhan, Hubei, China}
  \streetaddress{1037 Luoyu Road}
  \city{Wuhan}
  \state{Hubei}
  \country{China}
}

\author{Ming Liu}
\email{liuming@hust.edu.cn}
\affiliation{%
  \institution{School of Cyber Science and Engineering, Huazhong University of Science and Technology, Wuhan, Hubei, China}
  \streetaddress{1037 Luoyu Road}
  \city{Wuhan}
  \state{Hubei}
  \country{China}
}

\renewcommand{\shortauthors}{Ke Li et al.}

\begin{abstract}
  The ubiquitous adoption of Large Language Generation Models (LLMs) in programming has underscored the importance of differentiating between human-written code and code generated by intelligent models. This paper specifically aims to distinguish code generated by ChatGPT from that authored by humans. Our investigation reveals disparities in programming style, technical level, and readability between these two sources. Consequently, we develop a discriminative feature set for differentiation and evaluate its efficacy through ablation experiments. Additionally, we devise a dataset cleansing technique, which employs temporal and spatial segmentation, to mitigate the dearth of datasets and to secure high-caliber, uncontaminated datasets. To further enrich data resources, we employ "code transformation," "feature transformation," and "feature customization" techniques, generating an extensive dataset comprising 10,000 lines of ChatGPT-generated code. The salient contributions of our research include: proposing a discriminative feature set yielding high accuracy in differentiating ChatGPT-generated code from human-authored code in binary classification tasks; devising methods for generating extensive ChatGPT-generated codes; and introducing a dataset cleansing strategy that extracts immaculate, high-grade code datasets from open-source repositories, thus achieving exceptional accuracy in code authorship attribution tasks.
\end{abstract}

  


\maketitle
\section{Introduction}
Since its introduction in November 2022, OpenAI's ChatGPT has become the cynosure in numerous fields, generating palpable enthusiasm. By capitalizing on human feedback reinforcement learning (RLHF) for fine-tuning and judiciously curated datasets, ChatGPT exhibits exemplary capabilities across a plethora of challenging natural language processing (NLP) tasks. These include code synthesis via natural language\cite{r1}, text summarization\cite{r2}, and the creation of stylistic narratives based on designated elements\cite{r3}, in addition to its adeptness in conventional NLP tasks such as translation and text categorization. Moreover, ChatGPT demonstrates the prudence of refraining from responding to inquiries that exceed its knowledge base, contravene ethical norms, or broach sensitive political topics. \par
ChatGPT's prowess, particularly in facilitating programming through natural language, has engendered significant intrigue. However, reservations concerning the safety and legality of employing intelligent programming assistants like ChatGPT have been voiced by the scholarly community. The opaque nature of ChatGPT's training data spawns uncertainties regarding the provenance of its generated code snippets and the possibility of their harboring security vulnerabilities or unsafe code fragments\cite{r4, r16}, which could entangle developers in copyright disputes and code security quandaries.\par
For example, Stack Overflow, a renowned platform for programming inquiries, has imposed a temporary restraining on ChatGPT-generated content due to its low accuracy which did not match the high quality demanded by users. Concurrently, academic institutions face dilemmas in assessment settings where students might exploit ChatGPT to accomplish programming assignments, thus creating hurdles in evaluating their true acumen and knowledge.\par
Addressing these issues hinges on the adept discernment of human-written code versus code generated by intelligent models like ChatGPT. This entails two facets: first, the extraction of discriminative features between human-authored and ChatGPT-synthesized code. Conventional feature sets for code author attribution, which are primarily geared toward distinguishing among human authors, may fall short in capturing subtle distinctions, such as code structuring traits, when discerning between ChatGPT and human-authored code. This necessitates ablation studies to pinpoint a discriminative feature set capable of enhancing classification accuracy.\par
Through ablation studies, we identified variances in programming style, technical level, and readability between ChatGPT-generated and human-authored code. Building upon existing research in code authorship attribution, we formulate a discriminative feature set that enables distinction between these code sources, thereby surmounting feature analysis challenges.\par
The second facet concerns the compilation of an associated high-quality dataset. Our examination ascertained that prevailing research on code authorship attribution relies on datasets of heterogeneous quality and sources, bereft of standardized benchmarks. Moreover, datasets specifically comprising ChatGPT-generated code are scant owing to its recent advent.\par
To redress this dataset deficiency, we conceive a dataset cleansing technique grounded in temporal and spatial segmentation. This technique facilitates the procurement of untainted, high-quality datasets from open-source repositories while confirming data authenticity and eliminating extraneous factors related to authorship. Additionally, we employ a semi-automated approach to generate an expansive dataset, with ChatGPT synthesizing millions of lines of code via three strategies: "code transformation," "function transformation," and "custom functionality." This enriched dataset, which amalgamates human-authored and ChatGPT-generated code, furnishes robust data support for our research.\par

In summation, our research offers the following seminal contributions:
\begin{enumerate}
\item We devise a discriminative feature set predicated on heuristic code segmentation, which efficaciously differentiates ChatGPT-generated code from human-authored code, achieving a classification accuracy of over 90\% in code origin identification tasks through synergistic use with machine learning algorithms.
\item We introduce three strategies to amass a copious volume of ChatGPT-generated code: "code transformation," "function transformation," and "function customization."
\item We put forth a dataset cleansing technique centered on temporal and spatial segmentation, facilitating the extraction of immaculate, high-quality code datasets from open-source repositories. Employing this technique, we have amassed and purified approximately 20GB of code datasets. Subsequent application of these datasets to code author attribution tasks yielded an impressive accuracy rate of no less than 95\%. 

\end{enumerate}
\par
Our contributions establish a robust theoretical framework for distinguishing code generated by advanced models like ChatGPT from human-authored code. These findings bear considerable practical implications, including promoting academic integrity, protecting intellectual property, and bolstering software security, while fostering the sustainable development of AI in programming. This research is poised to catalyze advancements in related domains and pave the way for a harmonious coexistence between human developers and intelligent programming assistants.
\par
The remainder of this paper is organized as follows: Section 2 presents research work related to AI-generated content detection, code authorship attribution and ChatGPT-generated code. Section 3 provides an overview of the creation of datasets comprising human-generated code and ChatGPT-synthesized code. Section 4 elucidates the methodologies employed in developing the discriminative feature set. Section 5 details the experiments conducted and the ensuing analysis of the results. Section 6 concludes the paper and provides an outlook for further research. Finally, Section 7 outlines the limitations of our research and suggests potential directions for future research.

\section{Related Work}
\subsection{AIGC Detector}
\par
Research is already emerging in the field of detecting and identifying content generated by artificial intelligence (AI) models. For instance, in the domain of natural language processing, there have been notable studies focused on distinguishing AI-generated content from human-created content.\par
Mitrovi´c et al. \cite{r5} employed a machine learning approach to differentiate ChatGPT by extracting features from text messages. Their study achieved an accuracy of 79\% by focusing on shorter text responses generated by ChatGPT in comparison to manually generated text.\par
In another study, Guo et al. \cite{r6} collected 40K questions and corresponding answers from both human experts and ChatGPT to construct the Human ChatGPT Comparison Corpus (HC3) dataset. This dataset encompassed various domains, including open domain, finance, medical, legal, and psychological fields. Through the HC3 dataset, the researchers investigated differences between the features of ChatGPT-generated texts and those of human experts. They conducted feature analysis from multiple perspectives, including Vocabulary Features, Part-of-Speech and Dependency Analysis, and Sentiment Analysis. Additionally, they employed classical machine learning and deep learning techniques, performing experiments with three representative methods to further validate their findings.\par
Wenxiong Liao et al.\cite{r14} conducted a comprehensive analysis of medical texts to distinguish between content written by human experts and text generated by ChatGPT. Their study involved constructing a specialized dataset for medical texts and analyzing linguistic features, such as vocabulary, among others. To detect medical texts generated by ChatGPT, they employed a machine learning approach, specifically utilizing a BERT-based model. The results were promising, with the model achieving an impressive F1 score exceeding 95\%. Furthermore, the researchers observed that medical texts authored by humans tend to be more concrete, diverse, and contain a wealth of useful information. Conversely, medical texts generated by ChatGPT prioritize fluency and logic, often expressing general terminologies rather than providing context-specific, problem-related information.
\subsection{Code authorship attribution}
To date, there has been limited academic research addressing the specific problem of distinguishing code generated by AI models from code written by humans. However, this research can be aligned with the traditional code authorship attribution problem, which focuses on extracting code-level features to differentiate or trace the authorship of code.\par
In the field of code authorship attribution, classical machine learning approaches have been explored. Caliskan-Islam et al. \cite{r7} utilized a dataset comprising the code of 1600 authors from Google Code Jam. They extracted approximately 120,000 features based on vocabulary, layout, and syntax from the source code. By employing a random forest model consisting of 300 decision trees, they achieved up to 98\% accuracy on a test set comprising code from 250 programmers who participated in Google Code Jam 2014, effectively pushing the boundaries of machine learning models.\par
Regarding deep learning, Abuhamad et al. \cite{r8} employed a dataset consisting of 1600 authors from Google Code Jam (spanning from 2008 to 2016) and 1987 repositories from GitHub, with 142 C++ and 745 C programmers. They initially utilized the text analysis tool TF-IDF (Term Frequency-Inverse Document Frequency) for preprocessing the source code, which served as input for a deep learning network. They designed a deep learning framework based on recurrent neural networks to extract code features. The author attribution task was then accomplished using a random forest classifier. This approach achieved an accuracy of 96\% in the Google Code Jam experiment with 1600 authors and 94.38\% on the real dataset of 745 C programmers.
\subsection{Code generated by ChatGPT}
The current research on code generated by ChatGPT encompasses several areas, including code security, code correctness, code quality improvement, code error resolution, code meaning interpretation, and the potential of ChatGPT as a programming assistant. \par
Raphaël Khoury et al. \cite{r4} explored the capabilities of ChatGPT in generating programs and assessed the security of the resulting source code. They also investigated the effectiveness of prompting ChatGPT to enhance code security and delved into the ethical considerations associated with leveraging AI for code generation. The findings indicate that ChatGPT demonstrates some awareness of potential vulnerabilities. However, it frequently generates source code that lacks robustness against certain attacks.\par
Jiawei Liu et al. \cite{r16} introduced EvalPlus, a code synthesis benchmarking framework designed to thoroughly evaluate the functional correctness of code synthesized by Language Models (LLMs). In their work, they extended the widely used HUMANEVAL benchmark and created HUMANEVAL+, which includes an additional 81× generated tests. Through extensive evaluation across 14 popular LLMs, including GPT-4 and ChatGPT, they demonstrated that HUMANEVAL+ effectively detects a significant number of previously undetected erroneous code synthesized by LLMs. On average, it reduces the pass@k metric by 15.1\%.\par
The research conducted by Burak Yetiştiren et al.\cite{r15} aims to conduct a comparative analysis of prominent code generation tools, including GitHub Copilot, Amazon CodeWhisperer, and ChatGPT, in terms of various code quality metrics such as Code Validity, Code Correctness, Code Security, Code Reliability, and Code Maintainability. To achieve this, they utilize the benchmark HumanEval Dataset to evaluate the generated code based on the proposed code quality metrics.The analysis reveals that the latest versions of ChatGPT, GitHub Copilot, and Amazon CodeWhisperer achieve correct code generation rates of 65.2\%, 46.3\%, and 31.1\% respectively. Moreover, the newer versions of GitHub Copilot and Amazon CodeWhisperer demonstrate improvement rates of 18\% and 7\% respectively in terms of generating correct code. Additionally, the average technical debt, considering code smells, is found to be 8.9 minutes for ChatGPT, 9.1 minutes for GitHub Copilot, and 5.6 minutes for Amazon CodeWhisperer.\par
Jules White et al. \cite{r10}explores several prompt patterns that have been applied to improve requirements elicitation, rapid prototyping, code quality, refactoring, and system design.\par
Dominik Sobania et al. \cite{r9} evaluate the bug fixing performance of ChatGPT on the widely used QuixBugs benchmark set. They compare its performance with several other approaches reported in the literature. The findings reveal that ChatGPT's bug fixing capability is on par with common deep learning approaches like CoCoNut and Codex, and significantly outperforms standard program repair methods. Moreover, by providing hints to ChatGPT, they were able to further enhance its success rate, with ChatGPT successfully fixing 31 out of 40 bugs, surpassing the state-of-the-art performance.\par
Eason Chen et al.\cite{r11} presents GPTutor, a ChatGPT-powered rogramming tool, which is a Visual Studio Code extension using the ChatGPT API to provide programming code explanations. By integrating Visual Studio Code API, GPTutor can comprehensively analyze the provided code by referencing the relevant source codes. Preliminary evaluation indicates that GPTutor delivers the most concise and accurate explanations compared to vanilla ChatGPT and GitHub Copilot. Moreover, the feedback from students and teachers indicated that GPTutor is user-friendly and can explain given codes satisfactorily.\par
Haoye Tian et al. \cite{r12} conduct an empirical analysis to evaluate the capabilities of ChatGPT as a fully automated programming assistant, with a focus on code generation, program repair, and code summarization. The study specifically examines ChatGPT's performance in solving common programming problems and compares it to state-of-the-art approaches using two benchmark datasets. The findings demonstrate that ChatGPT effectively addresses typical programming challenges. However, the researchers also identify limitations in its attention span. They observe that when faced with comprehensive problem descriptions, ChatGPT's focus can be constrained, thus hindering its ability to leverage its vast knowledge for effective problem-solving.\par
Yihong Dong et al.\cite{r13} present a self-collaboration framework for code generation employing LLMs, exemplified by ChatGPT. Specifically, through role instructions, 1) Multiple LLMs act as distinct ``experts'', each responsible for a specific subtask within a complex task; 2) Specify the way to collaborate and interact, so that different roles form a virtual team to facilitate each other's work, ultimately the virtual team addresses code generation tasks collaboratively without the need for human intervention. They conduct comprehensive experiments on various code-generation benchmarks. Experimental results indicate that self-collaboration code generation relatively improves 29.9\%-47.1\% Pass@1 compared to direct code generation, achieving state-of-the-art performance and even surpassing GPT-4.

\section{Human and ChatGPT Code Dataset}
The dataset as a whole can be divided into two parts: those generated by ChatGPT and those crawled from Github.
\subsection{ChatGPT Code Dataset}
Our endeavor to comprehensively discern the attributes of code generated by ChatGPT necessitated the creation of corresponding datasets. We employed two versions of ChatGPT: GPT-3.5, which avails batch generation through an API, and GPT-4, which presently only facilitates generation through human interaction due to the absence of an open API.\par
It is imperative to recognize that ChatGPT's training data could encompass code from open-source repositories on Github. To investigate the viability of employing ChatGPT for code creation, and to discern its idiosyncratic features, we applied three distinct strategies: “code translation,” “function translation,” and “custom functions” for each version of ChatGPT. These strategies are explicated as follows:\par
\begin{itemize}
\item \textbf{Code Translation:} This strategy entails supplying ChatGPT with a code snippet written in programming language X, and tasking it to generate equivalent functionality code in a different programming language Y. The generated code must span at least 100 lines to circumvent the production of “boilerplate code” replete with annotations, which could impair data quality. As the generated code is translated into a different programming language, it is less likely to have been part of ChatGPT's training data, hence, it is relatively novel. But this also runs the risk of mimicking the wording and logic of the original code.
\item \textbf{Functional Translation:} In contrast to Code Translation, Functional Translation requires ChatGPT to first analyze and abstract a functional description from a given code snippet. Subsequently, ChatGPT is instructed to generate code in the same programming language as the original snippet based on the derived functional description. This method  prevents direct imitation of the original code, but it is slightly less innovative compared to Code Translation since the output is in the same language as the original code, which could be part of ChatGPT's training data.
\item \textbf{Functional customisation:} This strategy diverges from the prior two by prompting ChatGPT to generate code based on pre-existing functional descriptions sourced from programming competitions, textbooks, or other typical programming tasks. As this approach neither supplies ChatGPT with sample code nor restricts it to the original programming language, it engenders greater originality compared to Function Translation. Nonetheless, since solutions to these programming tasks might be accessible online and included in ChatGPT's training data, this approach is not as novel as Code Translation.

\end{itemize}
\subsection{Human Code Dataset}

Human-authored code datasets are integral to research in the domain of code authorship attribution. Present studies predominantly utilize datasets procured from four principal sources:
\begin{enumerate}

\item \textit{Educational Programming Tasks:} These datasets encompass code penned by students for class assignments. However, their restrictive scope and educational focus impede their generalizability.

\item \textit{Competitive Programming Archives (e.g., Google Code Jam):} Although these datasets are of high caliber, they might not accurately depict programming methodologies employed in real-world software development, owing to the specialized nature of competition problems and environments.

\item \textit{Open Source Repositories (e.g., Github):} Being the most reflective of real-world programming practices, these datasets are invaluable. However, ascertaining the sole authorship of the code is challenging due to collaborative projects and the opacity of development processes.

\item \textit{Textbook Supplements:} Code written by authors to supplement programming textbooks is also used. However, this data is limited by the scope of the textbooks and lacks generalization.

\end{enumerate}

Among these, datasets derived from open-source repositories are ostensibly the most authentic and efficacious for capturing coding styles. Github, being the world's largest open-source community, hosts numerous repositories. Although individuals frequently upload their code, the prevalence of shared or borrowed code is significant. Additionally, coding styles evolve over time, with discernible differences between code written at the nascent stages of learning and that written post-acquiring professional expertise. Consequently, indiscriminate code harvesting from repositories could compromise dataset quality and obfuscate the analysis of distinct coding styles. \par
To alleviate this issue, this paper introduces a methodology premised on temporal and spatial segmentation to collate code from individual repositories, thereby capturing the essence of an author's coding style during a specific timeframe. The ensuing section delineates the steps undertaken to clean the dataset.
\begin{enumerate}
\item \textit{Initial Cleanup:} In this step, organization accounts, forked repositories, and duplicates are eliminated. This culling is predicated on existing literature and conventions. The filtering of accounts is restricted to individual users, thereby excluding organization accounts. Forked repositories are omitted, as they typically do not contain original code. Moreover, duplicate repositories and files are removed to economize on storage and computation.
\item \textit{Temporal Segmentation:} Repositories are categorically segmented by their creation year. Studies\cite{r7, r8} suggest that an author's coding style remains relatively stable for approximately two years. Consequently, the dataset is divided based on the creation year, ensuring each segment contains code within a specific timeframe. In our experiments, repositories are classified from 2008 to 2022, rendering 15 categories. Subsequently, we focus on the code produced in the recent year (2021-2022).
\item \textit{Spatial Segmentation:} This step entails the removal of third-party libraries and collaborative repositories. Repositories containing third-party libraries are excluded based on naming conventions. Furthermore, repositories with multiple contributors are identified and removed by analyzing the "contributors" metadata. A comprehensive list of common third-party libraries for C++ and Java was compiled based on frequency analysis and practical development experience, serving as a filter criterion.
\end{enumerate}
In theory, post these cleaning steps, the residual source code should predominantly be authored by the individuals themselves. Nevertheless, some exceptions may include code obtained from external sources. Such instances are considered data noise and are disregarded.\par
The aforementioned steps ensure that the final dataset chiefly comprises code that reflects the individual authors' distinct coding styles.

\section{Discriminative Features Set}
In this study, we draw upon methodologies from traditional code authorship attribution; however, we tailor the feature extraction process to suit the distinct nature of our task. Traditional code authorship aims to discriminate between code written by different individuals, while our objective is to classify code written by humans as one category and code generated by ChatGPT as another. Consequently, this necessitates a modified approach to feature selection. We adapt the feature selection from traditional code authorship by conducting a heuristic code feature analysis. This enables us to construct a  discriminative feature set that effectively discerns between human-authored code and ChatGPT-generated code. This feature set comprises three main categories: lexical features, structural layout features, and semantic features. Notably, this refined feature set diverges from those traditionally employed in code authorship studies and is specifically tailored for our task. In the ensuing subsections, we expound upon the design methods for each category within the feature set.

\subsection{Lexical features}
Instead of analyzing the lexical features of the entire code, we segregate the vocabulary within the code into four distinct categories:

\begin{enumerate}
\item \textit{Comments and Strings:} This category includes single and multi-line comments, as well as strings enclosed in double quotes. These text blocks are indicative of the author's textual style.
\item \textit{Identifiers:} Comprising class names, method names, variable names, and interface names, identifiers reveal the author's naming conventions and library usage patterns.
\item \textit{Keywords:} These reserved words are intrinsic to the programming language, governing syntax structures, control flow, data types, and variable declarations. Analyzing the usage of keywords sheds light on the author's programming practices within the language.
\item \textit{Imported Libraries:} This category encompasses the standard and third-party libraries incorporated in the code through "include" (C++) or "import" (Java) statements. This reflects the author's familiarity with various libraries.
\end{enumerate}
Prior to lexical analysis, we tokenize the code, taking into account conventions such as camel case or underscores in identifiers. We separate words in comments, strings, and identifiers using spaces and punctuation marks. Subsequently, we split these tokens according to naming conventions and normalize them to lowercase. For keywords, we compare the tokens against a set of language-specific keywords. For imported libraries, we retain the complete names as they represent entities and are indicative of the author's style.\par
We tally the count of each vocabulary type and compute the term frequency (TF) of each word within these categories.

\subsection{Structural Layout Features}
In our preliminary analysis of ChatGPT code datasets, we observed that ChatGPT adheres to certain conventional formatting standards regarding layout features. While this is also typical of human-authored code, directly employing layout features from traditional code authorship as distinguishing factors would not be efficacious.\par
However, upon rigorous comparative analysis, we discerned subtle yet distinguishing layout and structural features that are characteristic of ChatGPT-generated code. We identified 22 such features, encompassing aspects such as comment ratio, blank line ratio, presence of line breaks preceding braces, average nesting depth, indentation length, and the average number of parameters in functions. These features are reflective of coding conventions and styles, and exhibit discernable disparities between human-authored code and ChatGPT-generated code. For a comprehensive listing of these features, refer to Table \ref{tab:features}.

\subsection{Semantic features}
In this study, we undertook an exhaustive, hands-on comparative analysis supplemented with extensive literature review to devise a set of semantic features intrinsic to code, an aspect that had remained untouched in prior research concerning code authorship attribution. It is our conviction that humans and AI embody distinct proficiencies and constraints when it comes to programming. For instance, human programmers excel in logical reasoning and possess the invaluable ability to collaborate and brainstorm, yet they are shackled by the boundaries of their knowledge and are susceptible to emotional biases such as procrastination or frustration. Contrarily, AI boasts an encyclopedic reservoir of knowledge, and remains impervious to emotional fluctuations; however, it falls short in logical reasoning prowess and lacks the initiative to scrutinize code autonomously. Interestingly, these disparities are mirrored in the semantics of the code written by humans and AI, and these semantics play a pivotal role in various facets like the code's execution efficiency, accuracy in real-world deployment, performance metrics, and so on. After meticulously evaluating various semantic features, we have zeroed in on the following three core aspects:
\begin{enumerate}
\item \textit{Runnability Analysis:} This entails compiling and executing the code to identify compilation or runtime errors.
\item \textit{Correctness Analysis:} If the code is runnable, we input test cases for algorithmic problems to verify whether the code produces accurate outputs.
\item \textit{Time-Space Performance Analysis:} For code that correctly solves the problem, we analyze the execution time and memory usage under large-scale test cases.
\end{enumerate}

\section{Differences between Code Authored by Humans and ChatGPT}
In Chapter 3, we presented the three categories of features that we identified as discriminative for distinguishing code produced by ChatGPT from human-authored code. In this chapter, we discuss experiments designed around these feature sets: binary classification, explanatory word frequency analysis, and exploratory analysis based on semantic features.

\subsection{Experiment Design}
\subsubsection{Binary Classification Experiment}
This experiment seeks to ascertain the feasibility of utilizing lexical and layout structural features to distinguish ChatGPT-generated code from human-written code, focusing on C++ and Java. We specifically employ lexical and layout structural features as they are readily quantifiable for machine learning models. For performance evaluation, we employ Accuracy, Precision, Recall, and F1 Score metrics, and also conduct ablation studies to investigate the contributions of each feature set.

\subsubsection{Explanatory Word Frequency Analysis Experiment}
This experiment visually represents and statistically analyzes the discrepancies in word usage within C++ and Java code authored by ChatGPT and humans. Specifically, we study the frequency of comments, strings, identifiers, keywords, and imported packages/headers. We contrast the frequencies and consider ChatGPT's documentation and relevant studies for additional context, offering analysis for particular variations.

\subsubsection{Exploratory Semantic Analysis}
\label{semantic_analysis}
This experiment investigates the semantic distinctions between code produced by ChatGPT and humans when solving identical programming problems. Due to the intricate nature of semantic feature extraction, and constraints in resources and time, this experiment mainly serves to provide insights and stimulate future research. Specifically, we present ChatGPT with 100 algorithm problems from LeetCode and evaluate various aspects, such as difficulty level, pass rate, executability, correctness, and time-space performance.

\subsection{Result Summarization}
\subsubsection{Effectiveness of Code Detection}
\begin{table*}[ht]
    \centering
    \small 
    \setlength{\tabcolsep}{4pt} 
    \caption{Comparison of Java code ablation experiment.}
    \label{tab:java-code-ablation}
    \begin{tabularx}{\textwidth}{l *{12}{c}} 
        \toprule
        & \multicolumn{4}{c}{lexical} & \multicolumn{4}{c}{layout} & \multicolumn{4}{c}{all} \\
        \cmidrule(lr){2-5} \cmidrule(lr){6-9} \cmidrule(lr){10-13}
        Algorithm & Accuracy & Precision & Recall & F-Measure & Accuracy & Precision & Recall & F-Measure & Accuracy & Precision & Recall & F-Measure \\
        \midrule
        Random Forest & 0.949 & 0.952 & 0.949 & 0.958 & 0.950 & 0.951 & 0.950 & 0.950 & 0.960 & 0.961 & 0.960 & 0.960 \\
        SMO & 0.940 & 0.940 & 0.940 & 0.940 & 0.878 & 0.883 & 0.878 & 0.877 & 0.958 & 0.959 & 0.958 & 0.958 \\
        Simple Logistic & 0.951 & 0.952 & 0.951 & 0.951 & 0.958 & 0.959 & 0.958 & 0.958 & 0.969 & 0.969 & 0.969 & 0.969 \\
        J48 & 0.974 & 0.974 & 0.974 & 0.974 & 0.949 & 0.950 & 0.949 & 0.949 & 0.978 & 0.978 & 0.978 & 0.978 \\
        \bottomrule
    \end{tabularx}
\end{table*}

\begin{table*}[ht]
    \centering
    \small 
    \setlength{\tabcolsep}{4pt} 
    \caption{Comparison of C++ code ablation experiments.}
    \label{tab:cpp-code-ablation}
    \begin{tabularx}{\textwidth}{l *{12}{c}} 
        \toprule
        & \multicolumn{4}{c}{lexical} & \multicolumn{4}{c}{layout} & \multicolumn{4}{c}{all} \\
        \cmidrule(lr){2-5} \cmidrule(lr){6-9} \cmidrule(lr){10-13}
        Algorithm & Accuracy & Precision & Recall & F-Measure & Accuracy & Precision & Recall & F-Measure & Accuracy & Precision & Recall & F-Measure \\
        \midrule
        Random Forest & 0.917 & 0.919 & 0.917 & 0.917 & 0.867 & 0.869 & 0.867 & 0.867 & 0.930 & 0.930 & 0.930 & 0.930 \\
        SMO & 0.900 & 0.902 & 0.900 & 0.900 & 0.863 & 0.864 & 0.863 & 0.863 & 0.924 & 0.927 & 0.924 & 0.924 \\
        Simple Logistic & 0.877 & 0.877 & 0.877 & 0.877 & 0.861 & 0.862 & 0.861 & 0.861 & 0.906 & 0.907 & 0.906 & 0.906 \\
        J48 & 0.884 & 0.886 & 0.884 & 0.884 & 0.796 & 0.796 & 0.796 & 0.796 & 0.890 & 0.892 & 0.890 & 0.890 \\
        \bottomrule
    \end{tabularx}
\end{table*}
\begin{figure}[ht]
  \centering
  \includegraphics[width=\linewidth]{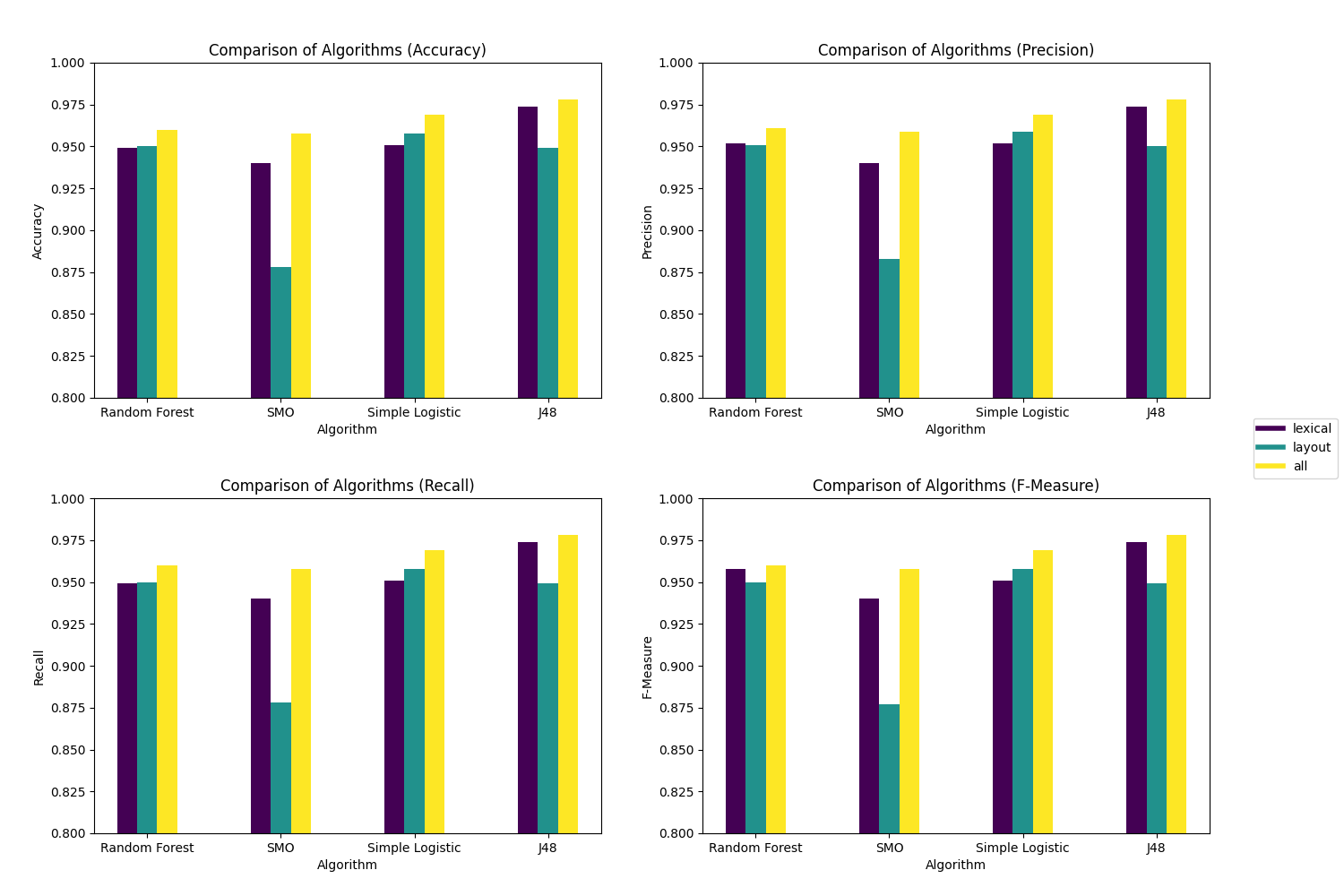}
  \caption{Comparison of of Java code ablation experiment.}
  \label{fig:java-ablation-experiment}
\end{figure}
\begin{figure}[ht]
  \centering
  \includegraphics[width=\linewidth]{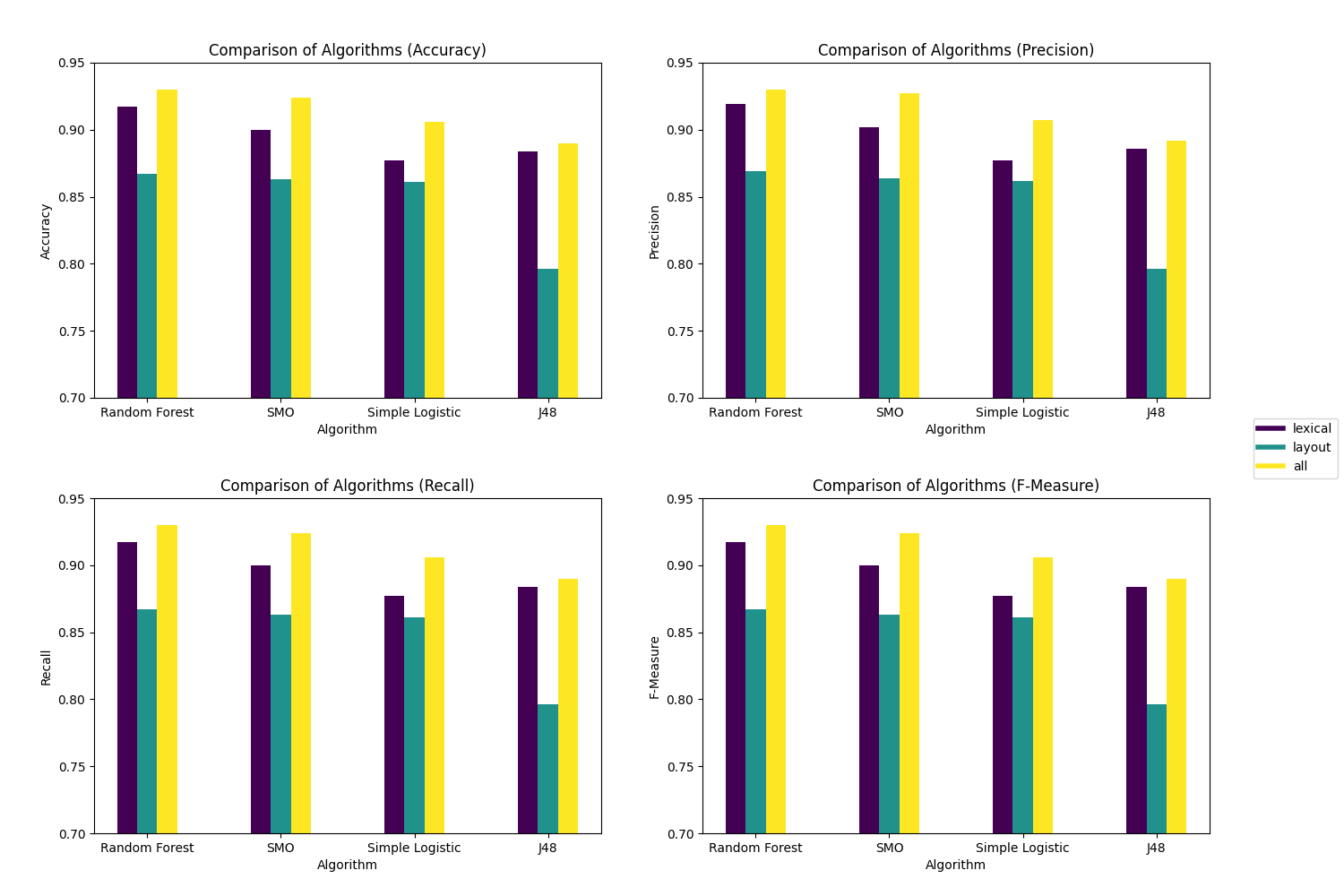}
  \caption{Comparison of C++ code ablation experiment.}
  \label{fig:cpp-ablation-experiment}
\end{figure}
\
\newline
\par
In the binary classification experiment, we analyzed the individual and combined effects of lexical and layout structural features in differentiating between human and ChatGPT-generated code. The results, as depicted in Table \ref{tab:java-code-ablation} and Table \ref{tab:cpp-code-ablation}, were visually presented in Figures \ref{fig:cpp-ablation-experiment} and \ref{fig:java-ablation-experiment}, respectively. These visual representations demonstrate that the combination of both feature sets yields the highest accuracy in classification. Notably, the complementary nature of these features in code classification tasks is underscored, implying the integral role each feature plays in distinguishing between the two code variants.
\subsubsection{Characteristic Differences}
\begin{figure}[ht]
  \centering
  \includegraphics[width=\linewidth]{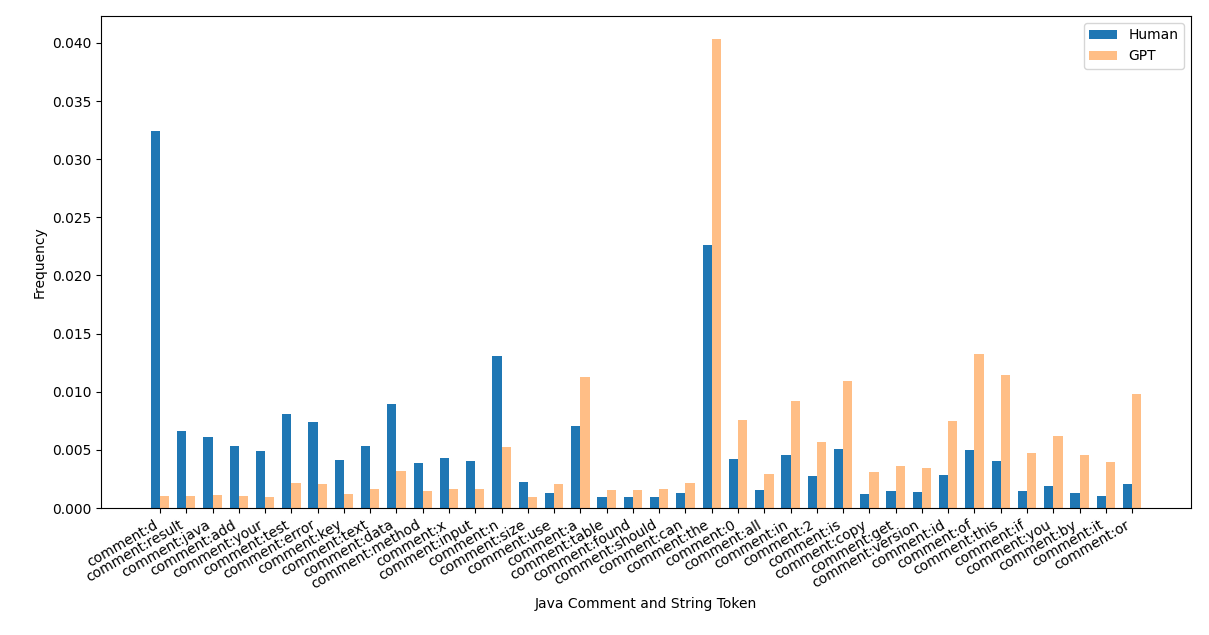}
  \caption{Comparison of comment and string frequencies in Java code.}
  \label{fig:java-exegesis-string}
  \Description{Comparison of comment and string frequencies in Java code.}
\end{figure}
\begin{figure}[ht]
  \centering
  \includegraphics[width=\linewidth]{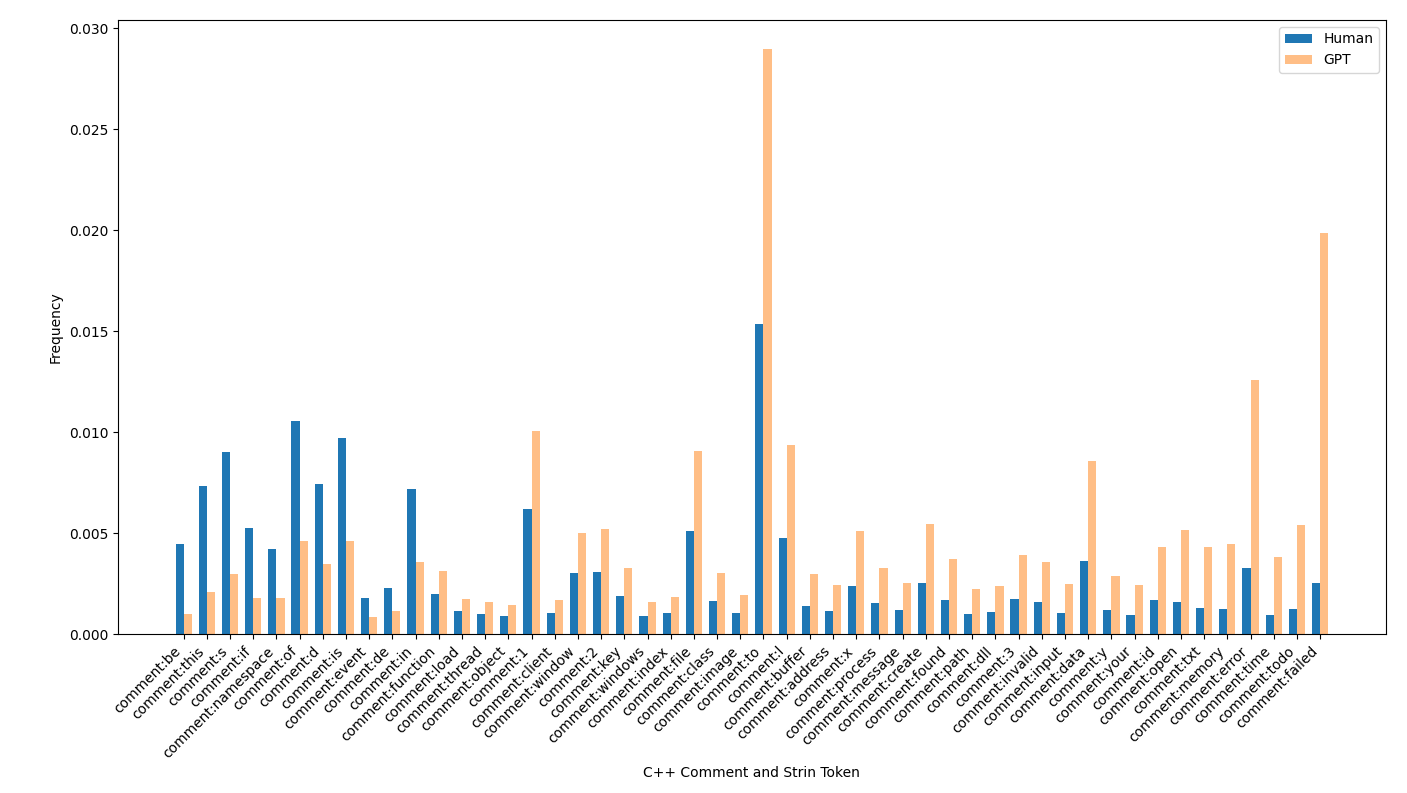}
  \caption{Comparison of comment and string frequencies in C++ code.}
  \label{fig:cpp-exegesis-string}
  \Description{Comparison of comment and string frequencies in C++ code.}
\end{figure}
\begin{figure}[ht]
  \centering
  \includegraphics[width=\linewidth]{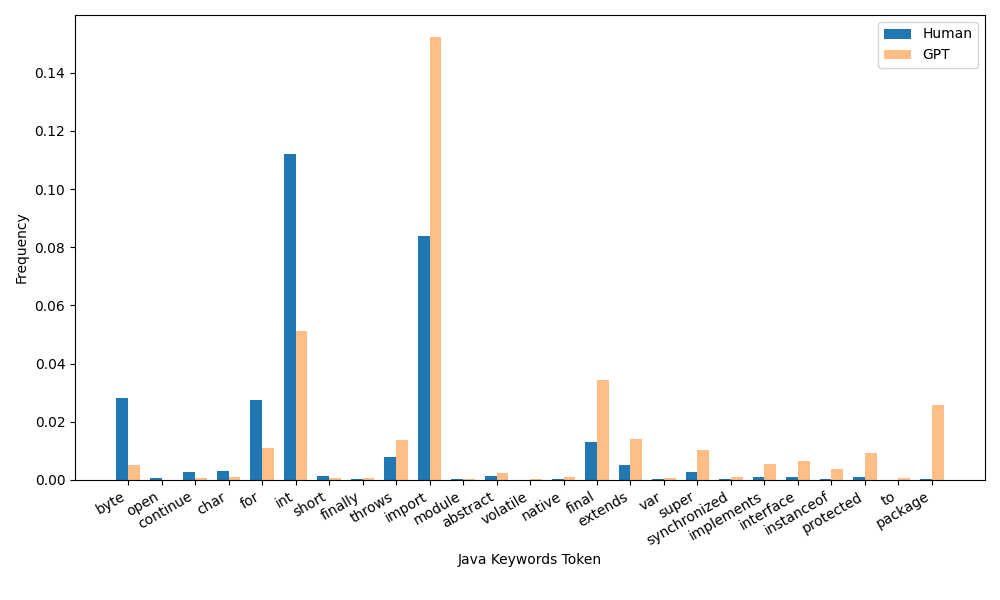}
  \caption{Comparison of keyword frequencies in Java code.}
  \label{fig:java-keywords}
  \Description{Java keywords}
\end{figure}
\begin{figure}[ht]
  \centering
  \includegraphics[width=\linewidth]{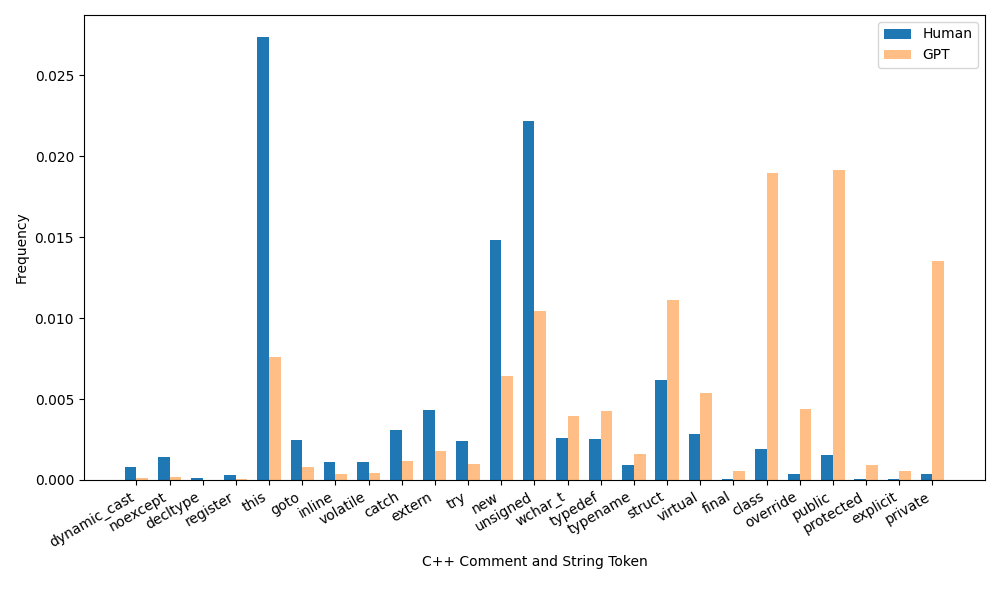}
  \caption{Comparison of keyword frequencies in C++ code.}
  \label{fig:cpp-keywords}
  \Description{C++ keywords}
\end{figure}
\begin{figure}[ht]
  \centering
  \includegraphics[width=\linewidth]{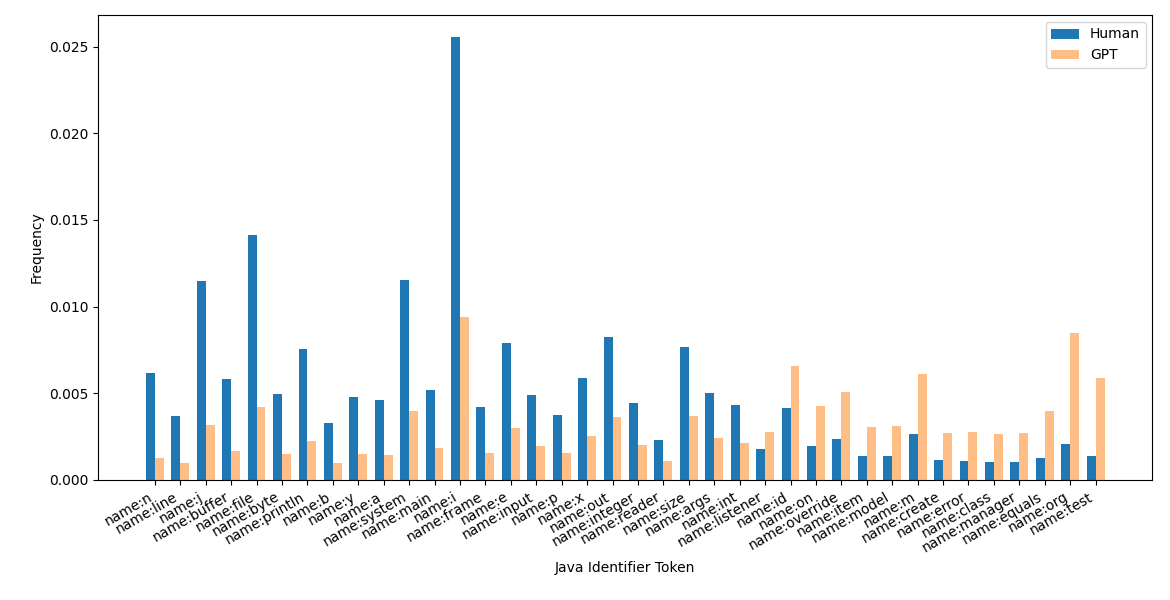}
  \caption{Comparison of identifier frequencies in Java code.}
  \label{fig:java-Identifiers}
  \Description{Java Identifiers}
\end{figure}
\begin{figure}[ht]
  \centering
  \includegraphics[width=\linewidth]{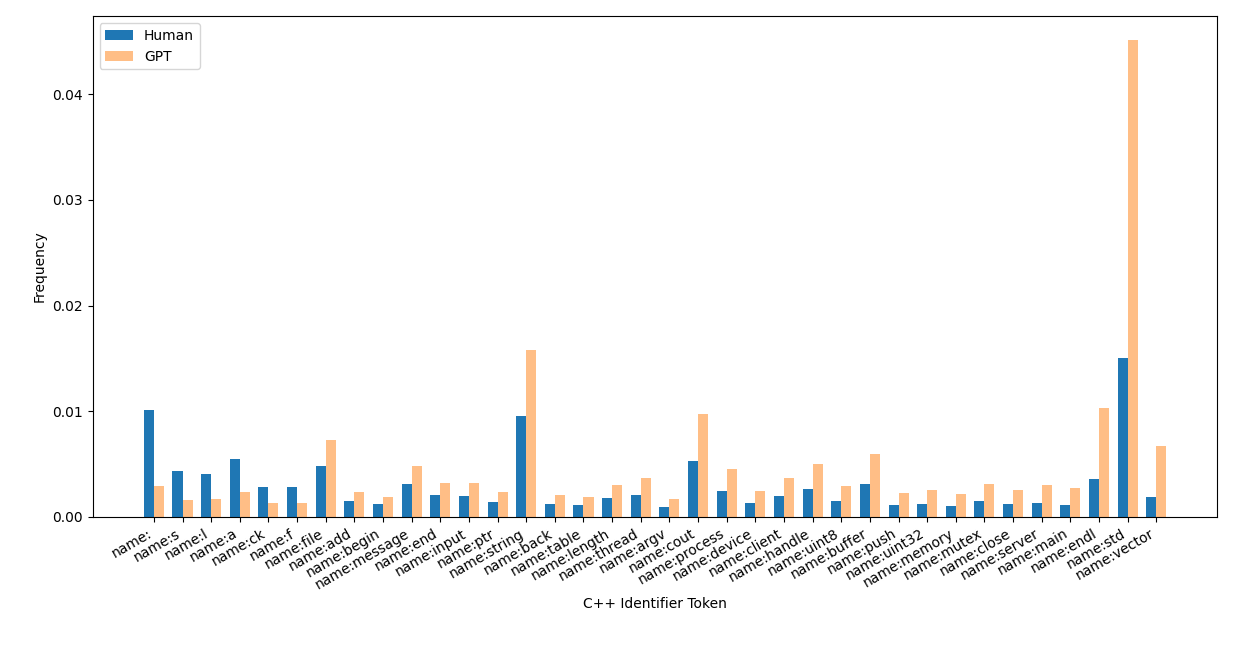}
  \caption{Comparison of identifier frequencies in C++ code.}
  \label{fig:cpp-Identifiers}
  \Description{C++ Identifiers}
\end{figure}
\begin{figure}[ht]
  \centering
  \includegraphics[width=\linewidth]{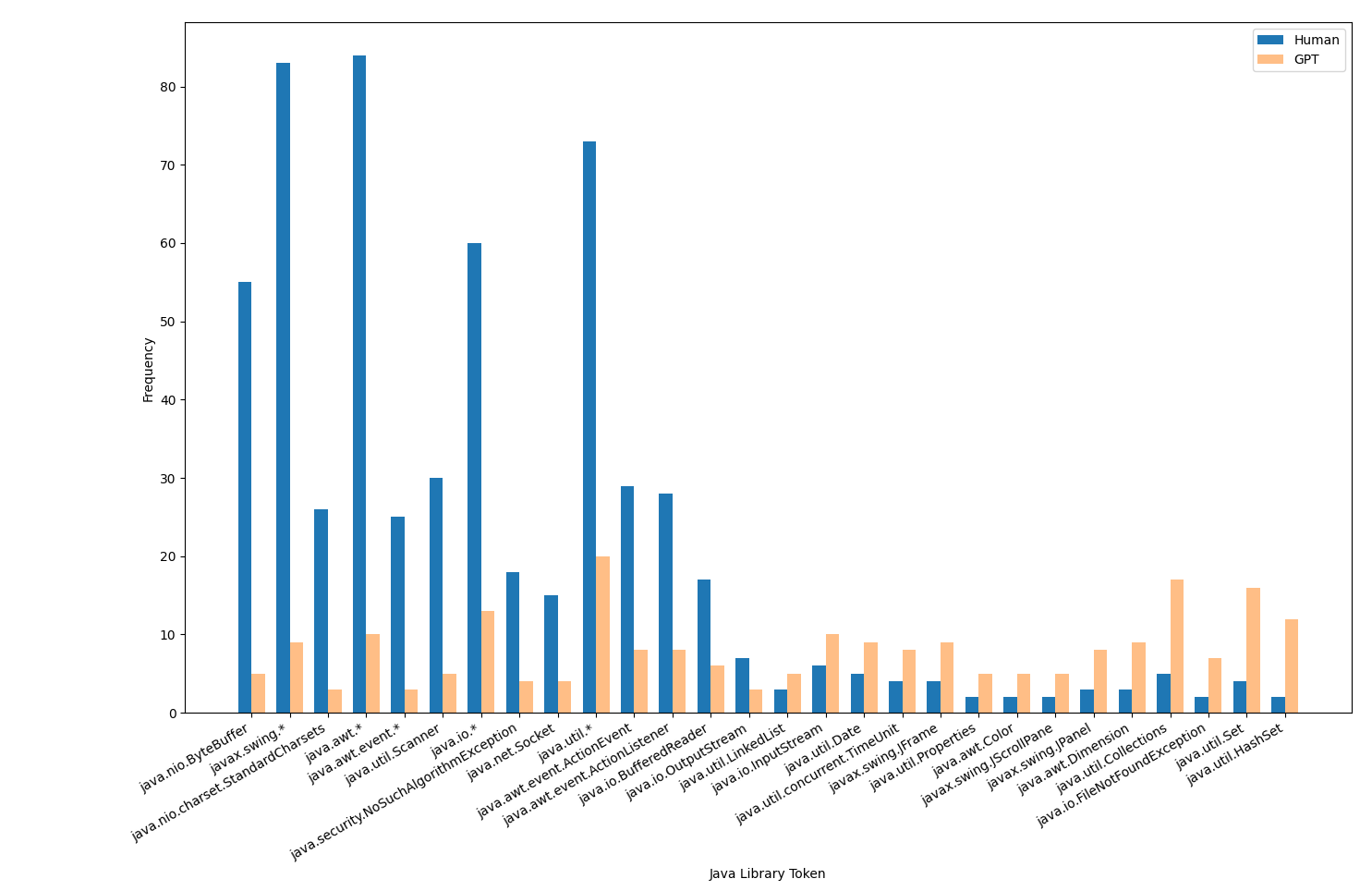}
    \caption{Comparison of library usage in Java code.}
    \label{fig:java-Third-party library}
  \Description{Java Third-party library}
\end{figure}
\begin{figure}[ht]
  \centering
  \includegraphics[width=\linewidth]{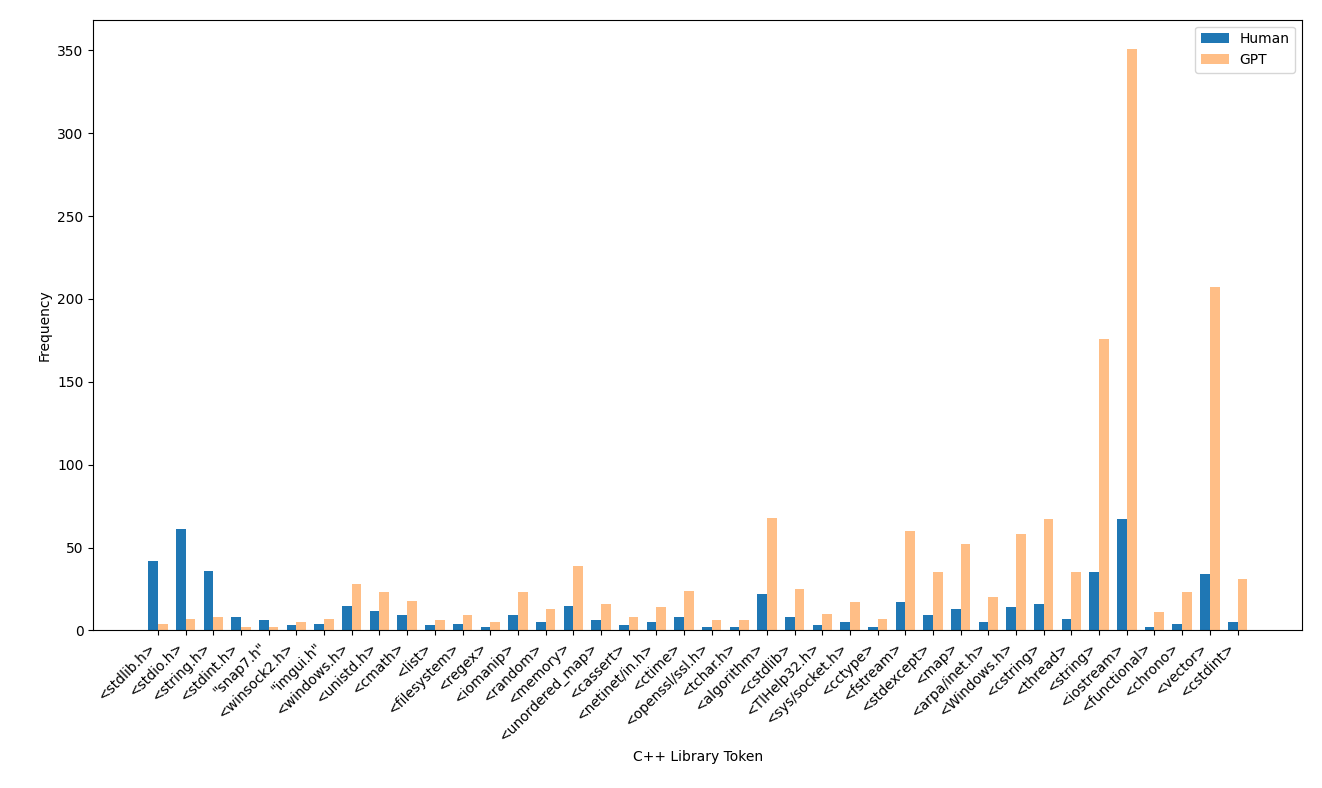}
  \caption{Comparison of library usage in C++ code.}
  \label{fig:cpp-Third-party library}
  \Description{C++ Third-party library}
\end{figure}
\
\newline
\par
In this subsection, we delve into the distinctions between ChatGPT-generated code and human-authored code by analyzing comments and strings, identifiers, keywords, and library usages. Figures \ref{fig:java-exegesis-string} to \ref{fig:cpp-Third-party library} depict these aspects, focusing on tokens with discrepancies exceeding 50\% and ranking them in descending order based on the difference. Our investigation aims to illuminate the inherent characteristics distinguishing ChatGPT-generated code from human-written code. The following observations can be drawn from the disparity in frequencies of these elements:
\begin{enumerate}
\item ChatGPT generally employs more elaborate and semantically rich terminology, whereas humans often favor concise and simpler words.
\item Human code frequently exhibits remnants of development iterations, such as commented-out code snippets, which are absent in ChatGPT-generated code. Additionally, ChatGPT does not generate extraneous code, while human code sometimes includes unreferenced variables or methods.
\item The code generated by ChatGPT adheres more closely to coding standards compared to human-authored code. For instance, in C++, ChatGPT prefers the standard “endl” line terminator over the “\textbackslash n” newline character.
\item ChatGPT-generated code typically does not exhibit inter-file dependencies, unlike human-written code, which often has a high degree of coupling between multiple files.
\item ChatGPT emphasizes readability, employing meaningful words for identifiers, whereas humans may use abbreviations or letters with ambiguous meanings.
\item ChatGPT is more likely to utilize newer language features. For example, in Java, it favors the “foreach” loop iteration over the traditional “for” loop, and in C++, it makes frequent use of the “auto” keyword for type inference.
\item ChatGPT-generated code resembles sample or template code sometimes, with comments often suggesting areas requiring custom implementation. In contrast, human code is typically intended for release, and employs logging for debugging instead of console output statements which are common in ChatGPT's code.
\end{enumerate}
\subsubsection{Examining the Originality and Semantic Similarities of ChatGPT-Generated Code}
\
\newline
\par
In our exploratory experiments, we sought to investigate if ChatGPT tends to replicate code authored by humans and to analyze the semantic differences between the code produced by humans and ChatGPT. To this end, we assembled a dataset composed of 100 programming tasks along with their respective human-authored solutions and codes. We then employed ChatGPT to generate code solutions for these tasks and performed a comparative analysis of the outcomes.\par
In the case of programming tasks predating ChatGPT's knowledge base cutoff date, the code generated by ChatGPT exhibited an exceptional executability rate of 100\% and a correctness rate of 100\%. In contrast, human solutions achieved an average correctness rate of 50\% under comparable circumstances. Additionally, ChatGPT's solutions demonstrated superior temporal efficiency in approximately 72.75\% of the cases and spatial efficiency in roughly 49.16\% of cases compared to human-authored code.\par
For programming tasks postdating ChatGPT's knowledge base cutoff date, ChatGPT achieved an executability rate of 90.9\%. The initial correctness rate of ChatGPT's solutions stood at 41\%, falling short of the human average of around 50\%. However, through subsequent refinements and iterations informed by follow-up questions, ChatGPT's solutions saw their correctness rate ascend to 63.63\%. Regarding temporal efficiency, ChatGPT's solutions outperformed human-authored code in approximately 58.91\% of cases. Similarly, in spatial efficiency, ChatGPT's solutions outstripped human-authored code in approximately 42\% of cases.\par
Furthermore, we conducted an in-depth comparison of ChatGPT-generated code against human-written code for the same programming tasks. Among the 100 code solutions, only two instances were observed in which ChatGPT exactly replicated human-authored code. In the majority of cases, ChatGPT manifested semantic similarity to human-authored code, echoing the underlying logic and ideas but employing distinct syntactic constructs and variations.\par
It is crucial to delve into the implications of ChatGPT's semantic similarity with human-authored code. The manifestation of semantic similarity suggests that ChatGPT is capable of discerning and emulating the core logic underlying human coding practices. While this could streamline the code generation process and uphold code quality, it also raises questions regarding innovation and originality in coding. ChatGPT's inclination to reflect existing coding conventions and logic could, in some instances, stifle novel approaches and solutions. Additionally, ethical considerations such as intellectual property rights and plagiarism need to be taken into account, especially in contexts where code is generated for commercial or proprietary applications. Ensuring that ChatGPT-generated code is sufficiently distinct from existing code, and that it acknowledges and credits foundational sources where necessary, is integral to the responsible and ethical deployment of this technology.

\section{Conclusion}
\par
This paper elucidates the critical distinctions between ChatGPT-generated code and human-authored code, along with their ramifications. The salient conclusions drawn from this investigation are enumerated below:
\begin{enumerate}
  \item The discriminative feature set proposed in this study demonstrates exceptional efficacy in discerning ChatGPT-generated code from human-authored code, with an accuracy rate surpassing 90\%. This underscores the viability of employing the identified feature set for distinguishing between ChatGPT-generated and human-authored code.
  \item Examination of token frequencies unveils pronounced disparities between ChatGPT-generated and human-authored code in the utilization of specific tokens. These discrepancies are indicative of disparate coding practices and preferences. Moreover, ChatGPT-generated code exhibits semantic variations compared to human-authored code when addressing identical programming tasks. These semantic differences represent varying levels of programming expertise and knowledge reservoirs.

  \item ChatGPT has demonstrated a propensity to reproduce existing code in scenarios where tasks are congruent with its pre-existing knowledge base. However, it is worth noting that this is not a simple replication; ChatGPT often recontextualizes and restructures the code, showcasing its ability to assimilate and adapt human programming expertise.

  \item While ChatGPT is proficient in generating functionally correct and efficient code, the semantic similarity with human-authored code raises questions regarding innovation and the ethical aspects of code generation. Ensuring originality and proper acknowledgment in ChatGPT-generated code is vital, especially in contexts involving proprietary or commercial applications.

\end{enumerate}
\par
The revelations from this study deepen our comprehension of the code generation capabilities of ChatGPT and bring into focus the parallels and distinctions between ChatGPT-generated and human-authored code. These insights are invaluable for diverse applications including code identification, analysis, and the integration of AI-assisted code generation into development workflows. Future inquiries can leverage these insights to bolster the precision and efficiency of code identification mechanisms and to investigate strategies for optimally combining the proficiencies of ChatGPT with human ingenuity. Moreover, additional studies could explore the ethical considerations and frameworks needed to guide the responsible use of AI in code generation.

\section{Limitations}
Notwithstanding the invaluable revelations procured from this investigation, it is imperative to recognize certain constraints that warrant consideration:
\begin{enumerate}
\item The ChatGPT code dataset used in this study is relatively small in size. Due to constraints in time and resources, the collected data is still insufficient and unbalanced across different sources, programming languages, styles, and tasks. To enhance the accuracy and reliability of ChatGPT code analysis and differentiation, a larger and more diverse dataset encompassing a wider range of coding styles and sources is needed.

\item The characterization of the ChatGPT code dataset is not exhaustive. It should be noted that all ChatGPT-generated code samples collected for this study were generated without any specific prompts or instructions. Therefore, the analysis and conclusions presented in this paper are based on the general programming style and state of ChatGPT. It is important to recognize that using specific prompts or instructions during code generation, such as excluding comments or employing a particular programming approach, may potentially interfere with our feature detection process and challenge the validity of the conclusions drawn in this study.

\item The ChatGPT code dataset might not fully represent the programming style of large-scale code generation models. Currently, ChatGPT is a generic large-scale language generation model that addresses various domains, including code generation. However, it is not exclusively designed as a dedicated large-scale code generation model. Therefore, the programming characteristics observed in the HCCD dataset may not entirely reflect the programming style and behavior of specialized large-scale code generation models.

\end{enumerate}
\par
These constraints delineate avenues for future inquiries and underscore the necessity for broader and more heterogeneous datasets, exhaustive examination of code generation directives, and exploration of the programming attributes of specialized large-scale code generation models. Redressing these constraints will catalyze a more encompassing understanding of code generation models and foster the evolution of more precise and dependable methodologies for differentiating between code engendered by artificial intelligence and that which is authored by humans. Additionally, ethical considerations surrounding code generation and intellectual property should also be integral components of future studies.

\begin{acks}
We express our heartfelt gratitude to Professor Fu Cai for generously providing the necessary datasets and computational resources that facilitated this study, whose invaluable guidance, support, and insights have been indispensable throughout the course of this research. We also acknowledge the tireless efforts and contributions of our fellow researchers and colleagues at Information Security Laboratory, School of Cyber Science and Engineering, Huazhong University of Science and Technology, especially Professor Wen Ming, Dr. Jiang Shuai, Li Wenke, M.A.\par
\end{acks}


\section*{Appendix}
This appendix contains a table that summarizes the features used in the analysis. Each feature is accompanied by a brief description to provide clarity on its role within the study. The table can be found on the following page.
\begin{table*}
  \caption{Summary of Features Used in Analysis}
  \begin{tabular}{p{5cm} p{10cm}}
      \toprule
      \textbf{Feature Name} & \textbf{Description} \\
      \midrule
      Control Structure Density & Logarithm of the ratio of the count of seven control structure-related keywords (do, else if, if, else, switch, for, while) to file length \\
      Ternary Operator Density & Logarithm of the ratio of the count of ternary operators to file length \\
      Token Density & Logarithm of the ratio of the count of tokens to file length \\
      Comment Density & Logarithm of the ratio of the count of comments to file length \\
      Literal Density & Logarithm of the ratio of the count of literals to file length \\
      Keyword Density & Logarithm of the ratio of the count of keywords to file length \\
      Function Density & Logarithm of the ratio of the count of functions to file length \\
      Maximum Nesting Depth & Maximum nesting depth of control and loop structures \\
      Average Branching Factor & Average number of subtrees per code block \\
      Average Parameters per Function & Average number of parameters in functions \\
      Standard Deviation of Parameter Count & Standard deviation of parameter count in functions \\
      Average Line Length & Average length of lines in the code file \\
      Line Length Standard Deviation & Standard deviation of line lengths in the code file \\
      Macro Density & Logarithm of the ratio of the count of preprocessor macros to file length \\
      Tab Character Density & Logarithm of the ratio of the count of tab characters to file length \\
      Space Character Density & Logarithm of the ratio of the count of space characters to file length \\
      Empty Line Density & Logarithm of the ratio of the count of empty lines to file length \\
      Whitespace Ratio & Ratio of whitespace characters (spaces, tabs, new lines) to non-whitespace characters \\
      New Line Preceding Open Brace & Presence of a new line character before opening braces in code blocks \\
      Leading Indentation Type & Indentation at the beginning of each line using tabs or spaces \\
      Maximum AST Node Depth & Maximum depth of nodes in the Abstract Syntax Tree (AST) \\
      AST Node Bigram Frequencies & Relative frequencies of AST node bigrams \\
      Average AST Node Type Depth & Average depth of nodes of each type in the AST \\
      Keyword Frequencies & Relative frequencies of keywords in the code \\
      Average Code Depth in AST Leaves & Average depth of code in AST leaf nodes \\
      Line Length Frequencies & Frequencies of different line lengths in the code file \\
      \bottomrule
  \end{tabular}
  \label{tab:features}
\end{table*}
\end{document}